\documentclass[prx,twocolumn]{revtex4}
\usepackage{hyperref}
\usepackage{amssymb}
\usepackage{graphicx}
\usepackage{color}

\newcommand{\avg}[1]{\left\langle {#1} \right\rangle}

\newcommand{\comm}[2]{\left[ {#1}, {#2} \right]}

\newcommand{\beq}{\begin{equation}}
\newcommand{\eeq}{\end{equation}}

\begin{document}

\title{Coherent controllers for optical-feedback cooling of quantum oscillators}

\author{Ryan Hamerly}\email{rhamerly@stanford.edu}
\author{Hideo Mabuchi}\email{hmabuchi@stanford.edu}
\affiliation{Edward L.\ Ginzton Laboratory, Stanford University, Stanford, CA 94305}
	
\date{\today}
	
\begin{abstract}
We study the cooling performance of optical-feedback controllers for open optical and mechanical resonators in the Linear Quadratic Gaussian setting of stochastic control theory. We utilize analysis and numerical optimization of closed-loop models based on quantum stochastic differential equations to show that coherent control schemes, where we embed the resonator in an interferometer to achieve all-optical feedback, can outperform optimal measurement-based feedback control schemes in the quantum regime of low steady-state excitation number. These  performance gains are attributed to the coherent controller's ability to simultaneously process both quadratures of an optical probe field without measurement or loss of fidelity, and may guide the design of coherent feedback schemes for more general problems of robust nonlinear and robust control.
\end{abstract}

\maketitle	
	
As present-day engineering relies broadly and implicitly on real-time feedback control methodology~\cite{AstromMurray}, it is difficult to imagine our nascent explorations of quantum engineering advancing to technological relevance without rigorous extensions of core control theory to incorporate novel features of quantum dynamics, stochastics and measurement. While significant progress has been made recently in terms of analyzing quantum feedback systems~\cite{Bela83,Wise93,Dohe00,Mabu05,Dong10,Brif10} and in experimental demonstrations of quantum feedback control~\cite{Smit02,Arme02,Bush06,Giga06,Arci06,Klec06,Mabu08,Gill10,Sayr11,Iida11}, we still have a relatively limited understanding of systematic approaches to quantum control design and of the qualitative role of quantum coherence and entanglement between the plant and controller in a feedback loop.

Within the elementary context of linear open quantum systems, James, Nurdin and Petersen~\cite{JamesNurdinPetersen08,NurdinLQG} have utilized interconnection models based quantum stochastic differential equations (QSDEs)~\cite{Huds84,Carm93,Gard93,Barc06} to develop quantum generalizations of the traditional paradigms of ${\cal H}^\infty$ and Linear Quadratic Gaussian (LQG) optimal control. While some of the most exciting potential applications of quantum feedback control involve nonlinear dynamics and/or non-Gaussian noises~\cite{BitFlipPaper,BaconShorPaper,Mabu11a,Goug11}, the linear setting is an essential starting point for rigorous study and presents crucial advantages in terms of analytic and computational tractability.

Here we focus on a theoretical investigation of steady-state cooling of open quantum oscillators such as optical and optomechanical resonators subject to stationary heating, damping, and optical probing and feedback. We work within an LQG framework as in the recent paper of Nurdin, James and Petersen~\cite{NurdinLQG} and utilize numerical optimization together with fundamental analytic results~\cite{AstromMurray} bounding the best possible LQG performance of measurement-based feedback control schemes to establish and to interpret quantitative advantages of coherent feedback for cooling-type performance metrics in certain parameter regimes.

Following recent convention, as in~\cite{JamesNurdinPetersen08,NurdinLQG,Mabu08}, we will here refer to measurement-based controllers as ``classical'' controllers and to coherent feedback controllers as ``quantum'' controllers. This terminology reflects the general distinction that the signal processing required to determine LQG-optimal control actions from a real-time measurement signal can be implemented by a classical electric circuit, while all of the hardware in a coherent feedback loop must be physically describable using quantum mechanics (typically with weak damping).

\section{Linear Systems}

Quantum harmonic oscillators can be modeled as cascadable open quantum systems using the SLH framework \cite{Gough08,Gough09} and the associated QSDEs. In the SLH framework, any open quantum system may be described as a triple:
\beq
	G = (S, L, H)
\eeq
where $S$ is a scattering matrix, $L$ is a coupling vector and $H$ is the Hamiltonian operator for the system's internal degrees of freedom. For a {\em linear} system with an internal state $x$, $S_{ij}$ is independent of the internal state, $L_i = \Lambda_i x + \lambda_i$ is at most linear, and $H = \frac{1}{2} x^{\rm T} R x + r^{\rm T} x$ is at most quadratic.

Armed with an SLH representation the most efficient way to simulate a {\it linear} quantum system is to solve the QSDEs, which represent coupled Heisenberg equations of motion for system operators and input-output quantum stochastic processes. Following the work of James, Nurdin and Petersen~\cite{JamesNurdinPetersen08} we write the QSDEs for a linear system in the state-space form,

\begin{eqnarray}
	dx(t) & = & \left[A\,x(t) + a\right] dt + B\,da(t) \nonumber \\
	d\tilde{a}(t) & = & \left[C\,x(t) + c\right] dt + D\,da(t) \label{eq:abcd}
\end{eqnarray}

Here $x(t)$ gives the plant's internal variables; this is a Hermitian, operator-valued vector. $A$, $B$, $C$ and $D$ are real matrices; $a$ and $c$ are real vectors.  The processes $da(t)$ and $d\tilde{a}(t)$ are quantum stochastic processes for the inputs and outputs, respectively. For convenience, we make them Hermitian as well; for a given port, one has $da_i = \bigl(dA_i + dA_i^\dagger, (dA_i - dA_i^\dagger)/i\bigr)$, where $dA(t)$ is the quantum Wiener process~\cite{Gard04,Bout07} following the It\^{o} rule $dA_i\,dA_j^\dagger = \delta_{ij} dt$.

Defining $\Theta_{ij} = [x_i, x_j]/2i$ as the commutator matrix, the ABCD parameters of (\ref{eq:abcd}) can be related to the SLH parameters as follows:

\beq
\begin{array}{rclrcl}
	A & = & 2\Theta\left(R + \frac{1}{4}\tilde{\Lambda}^{\rm T} J \tilde{\Lambda}\right),\ \ \  &
	B & = & \Theta \tilde{\Lambda}^{\rm T} J \tilde{S}, \\
	C & = & \tilde{\Lambda}, &
	D & = & \tilde{S}, \\
	a & = & 2\Theta\left(r + \frac{1}{4}\tilde{\Lambda}^{\rm T} J \tilde{\lambda}\right), &
	c & = & \tilde{\lambda}
\end{array}
\eeq

(Here $\tilde{S}$, $\tilde{\Lambda}$, and $\tilde{\lambda}$ are real matrices which can be easily constructed from $S$, $\Lambda$ and $\lambda$, which are in general complex.  $J$ is a canonical antisymmetric matrix of the appropriate size.  See Appendix \ref{sec:slh-app}.)

To measure the performance of a given controller we need to define a cost function.  For example, to minimize the plant's response to a noisy input one could minimize the steady-state expectation value of the excitation number $\avg{a^\dagger a}$. With (classical) state feedback and in the absence of exogenous noise such a quadratic cost function would result in a Linear Quadratic Regulator (LQR) optimal control problem~\cite{AstromMurray}, but in our optical feedback scenario with Gaussian input fields (vacuum or thermal noise) this becomes a quantum LQG problem~\cite{AstromMurray,NurdinLQG}.

It is straightforward to concatenate and cascade linear systems once we have the ABCD models. We have written software in {\it Mathematica} to compute the ABCD matrices for an arbitrary linear quantum system~\footnote{Please contact the authors for distribution of the code.}.  This borrows many elements from the Modelica quantum circuit toolkit of Sarma et al.\ \cite{QHDLM}, and is similar to the QHDL framework of Tezak et al.\ \cite{QHDLPaper}.  The code computes the LQR cost function as a function of the plant and controller properties, and it would not be difficult to extend it to more general cost functions.  Thanks to the linearity of our system, simulation is very fast: the complexity is polynomial in the size of the circuit, not exponential as is usually the case for quantum simulations, and for a simple system, it computes the LQR in well under 50 microseconds.

Given a particular plant, the code is fast enough to perform a multivariate Newton-Raphson optimization scheme to find the (locally) optimal controller parameters.  This is possible regardless of whether the controller has any particular structure -- if the controller's structure is left arbitrary, the code can simply optimize with respect to the controller's ABCD matrices, subject to the physical realizability conditions
\begin{eqnarray}
	A \Theta + \Theta A^{\rm T} + B J B^{\rm T} & = & 0 \nonumber \\
	\Theta C^{\rm T} + B J D^{\rm T} & = & 0 \nonumber \\
	D J D^{\rm T} & = & J \label{eq:prc}
\end{eqnarray}
that arise from the fact that time evolution should preserve the commutation relations between system and input/output fields \cite{JamesNurdinPetersen08, NurdinLQG}.  Optimizing with respect to an ``arbitrary'' controller takes longer because there are more free parameters, but the code is fast enough for each Newton step to take no more than 1.5 milliseconds on a standard laptop.

We note that the classical steady-state LQG problem is a convex problem, and the optimal steady-state controller parameters can be derived via solution of algebraic Riccati equations~\cite{AstromMurray}. In the quantum case, no such closed-from solutions are known and the realizability constraints (\ref{eq:prc}) make the landscape for numerical optimization non-convex~\cite{NurdinLQG}. Hence while we can be sure about the classical optimality of measurement-based controllers for the oscillator cooling scenarios we consider, the coherent controllers we find via numerical optimization are merely {\it local} minima and can only be considered as candidates for quantum optimality.

\section{Control of an Optical Cavity}

\begin{figure}[t]
	\centering
	\includegraphics[width=1.00\columnwidth]{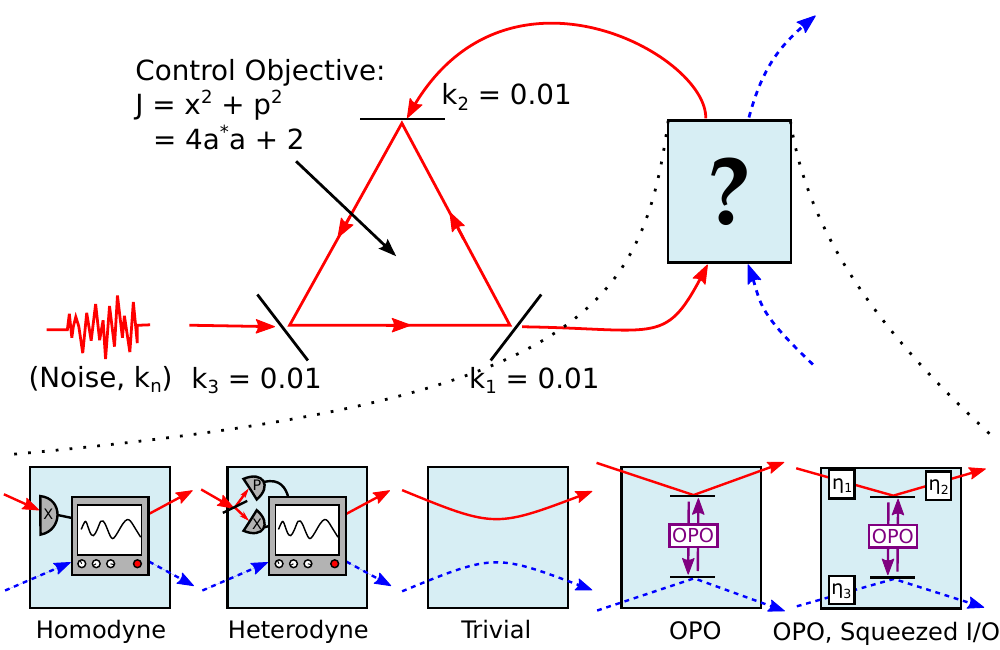}
	\caption{Optical cavity plant system with five possible classical and coherent feedback controllers}
	\label{fig:f1}
\end{figure}

As a simple example of a quantum ``plant'' system, consider an optical cavity with a noisy input, Fig.\ \ref{fig:f1}.  In the controller's absence, the cavity is driven by two vacuum inputs (mirrors $k_1$ and $k_2$, and one thermal input (mirror $k_3$).  Any noise process that is much broader spectrally than the cavity linewidth can be approximated as a ``white noise'' thermal input.  Without such noise, the cavity's internal mode decays quickly to the ground state.  The objective in this control problem is to minimize the effect of the noise on the cavity's internal state -- in other words, to minimize the photon number $\avg{a^\dagger a}$ of the cavity.  We accomplish this by sending output $1$ through a control circuit and feeding the result back into input $2$.  This is an LQG feedback control problem.

Five possible controllers are shown in Figure \ref{fig:f1}.  The classical controllers work by measuring a quadrature from the cavity's output (or in the heterodyne case, splitting the beam and measuring two different quadratures), and applying a feedback signal based on this measurement and the controller's internal state.  The ``trivial controller'' works by feeding the output directly back into mirror 2 of the plant, perhaps with a phase shift.  If the light reflecting off of mirror 2 is in phase with the light leaking out of the mirror, the light lost through both mirrors interferes constructively, reducing the control objective $\avg{a^\dagger a}$ (see also~\cite{Mabu11a}).

\begin{figure}[t!]
	\centering
	\includegraphics[width=1.00\columnwidth]{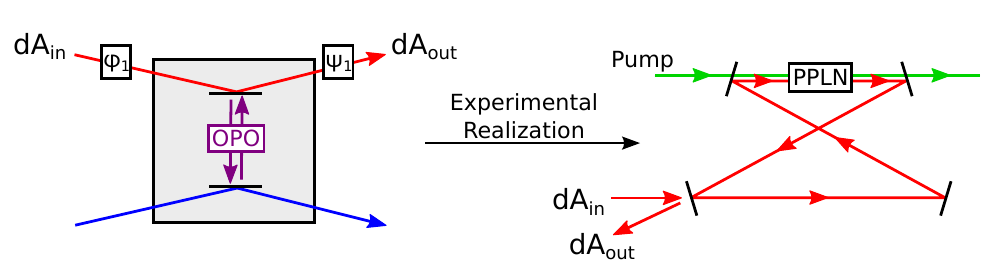}
	\caption{Experimental realization of an OPO with a cavity and a nonlinear crystal.}
	\label{fig:f2}
\end{figure}

The remaining two controllers shown in the figure are coherent controllers with memory.  Unlike the trivial controller, the control signal is a function not only of the input field, but also the input's history.  But unlike the classical controllers, the input field is not measured; instead, it is coherently processed and the result is fed back into the plant cavity.  These designs use an optical parametric oscillator (OPO, as in Fig.~\ref{fig:f2}) to squeeze the optical field.

The OPO will have the following SLH model:
\begin{eqnarray}
	& & S = 1_{2\times 2},\ \ L = \left[\sqrt{\kappa_1} a,\ \sqrt{\kappa_2} a\right], \nonumber \\
	& & H = \frac{1}{4} x^{\rm T} \left[\begin{array}{cc} \Delta - \mbox{Im}(\epsilon) & \mbox{Re}(\epsilon) \\ \mbox{Re}(\epsilon) & \Delta + \mbox{Im}(\epsilon) \end{array}\right]x
	\label{eq:opo-slh}
\end{eqnarray}
Here, $\kappa_1$ is related to the input/output mirror reflectance, $\kappa_2$ to other losses; $\Delta$ is the cavity detuning, and $\epsilon$ is a complex number, whose amplitude encodes the strength of the pump field and the nonlinear medium's $\chi^{(2)}$, and whose phase encodes the pump field's phase \cite{NurdinNetworkSynthesis}.

The plant system, an optical cavity with a noisy input, can be modeled as an open quantum system with three couplings, one for each mirror.  The SLH model for this system is:
\beq
	S = 1_{3\times 3},\ \ L = \left[\sqrt{k_1} a,\ \sqrt{k_2} a,\ \sqrt{k_3} a\right],\ \
	 H = \Delta a^\dagger a
\eeq
We also need to find the covariance matrix $F_{ij}$ for the noisy inputs $da_i$, defined  by $\frac{1}{2}\avg{da_i da_j + da_j da_i} = F_{ij}dt$.  Recall that, for {\it vacuum} inputs, the fields $dA$ and $dA^\dagger$ satisfy the It\^{o} relations $dA\,dA = dA^\dagger dA^\dagger = dA^\dagger dA = 0$, $dA\,dA^\dagger = dt$ \cite{Huds84,Bout07}, leading to the It\^{o} tables:

\begin{table}[h!]
	\centering
	\begin{tabular}{ccc}
	
	\begin{tabular}{c|cc}
		$dX$/$dY$ & $dA$ & $dA^\dagger$ \\ \hline
		$dA$ & $0$ & $dt$ \\
		$dA^\dagger$ & $0$ & $0$
	\end{tabular}
	&\ \ \ $\leftrightarrow$\ \ \ &
	\begin{tabular}{c|cc}
		$dx$/$dy$ & $da_x$ & $da_p$ \\ \hline
		$da_x$ & $dt$ & $i\,dt$ \\
		$da_p$ & $-i\,dt$ & $dt$
	\end{tabular}

	\end{tabular}	
\end{table}

For a {\it non-vacuum}, thermal input, the field $dA$ has additional (unsqueezed) noise, so $dA^\dagger dA = k_n dt$ for some noise strength $k_n > 0$, and the rest of the relations are adjusted accordingly, leading to the following It\^{o} tables:
\begin{table}[h!]
	\centering
	\begin{tabular}{ccc}
	
	\begin{tabular}{c|cc}
		$dX$/$dY$ & $dA$ & $dA^\dagger$ \\ \hline
		$dA$ & $0$ & $\!\!(1 + k_n)dt$ \\
		$dA^\dagger\!\!$ & $k_n dt$ & $0$
	\end{tabular}
	&\ \ $\leftrightarrow$\ \ &
	\begin{tabular}{c|cc}
		$dx$/$dy$ & $da_x$ & $da_p$ \\ \hline
		$da_x$ & $(1+2k_n)dt\!\!\!$ & $i\,dt$ \\
		$da_p$ & $-i\,dt$ & $\!\!\!(1+2k_n)dt$
	\end{tabular}

	\end{tabular}	
\end{table}

\begin{figure}[t]
	\centering
	\includegraphics[width=1.00\columnwidth]{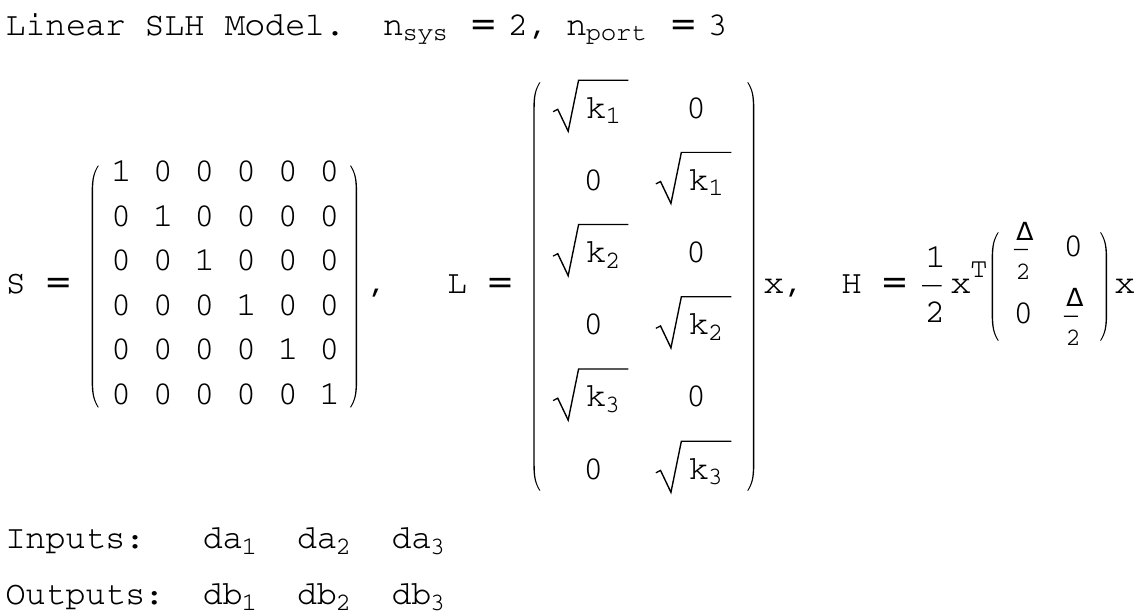}
	\caption{Output from our {\it Mathematica} package, describing the plant system.}
	\label{fig:f3}
\end{figure}

\noindent In the present system, inputs $1$ and $2$ are vacuum, and $3$ is thermal noise.  This gives the following covariance matrix:
\beq
	F = \left[\begin{array}{cccccc}
		1 & 0 & 0 & 0 & 0 & 0 \\
		0 & 1 & 0 & 0 & 0 & 0 \\
		0 & 0 & 1 & 0 & 0 & 0 \\
		0 & 0 & 0 & 1 & 0 & 0 \\
		0 & 0 & 0 & 0 & 1+2k_n & 0 \\
		0 & 0 & 0 & 0 & 0 & 1+2k_n
		\end{array}\right]
\eeq
The plant system is easy to set up in our {\it Mathematica} package; a sample output is shown in Figure \ref{fig:f3}.  The package, based on the circuit modeling and analysis framework of Sarma et al.\ \cite{QHDLM}, allows one to arbitrarily concatenate and link smaller elements to form larger quantum circuits, as long as all of the components are linear.  The feedback control circuit is one example system the package can be used to simulate.

Once the combined plant / controller system is set up, with its associated $A$, $B$, $C$ and $D$ matrices, the covariance matrix $\sigma_{ij} = \frac{1}{2} \avg{x_i x_j + x_j x_i}$ can be computed with the Lyapunov equation
\beq
	A \sigma + \sigma A^{\rm T} + B F B^{\rm T} = 0
\eeq

\begin{figure}[b]
	\centering
	\includegraphics[width=1.00\columnwidth]{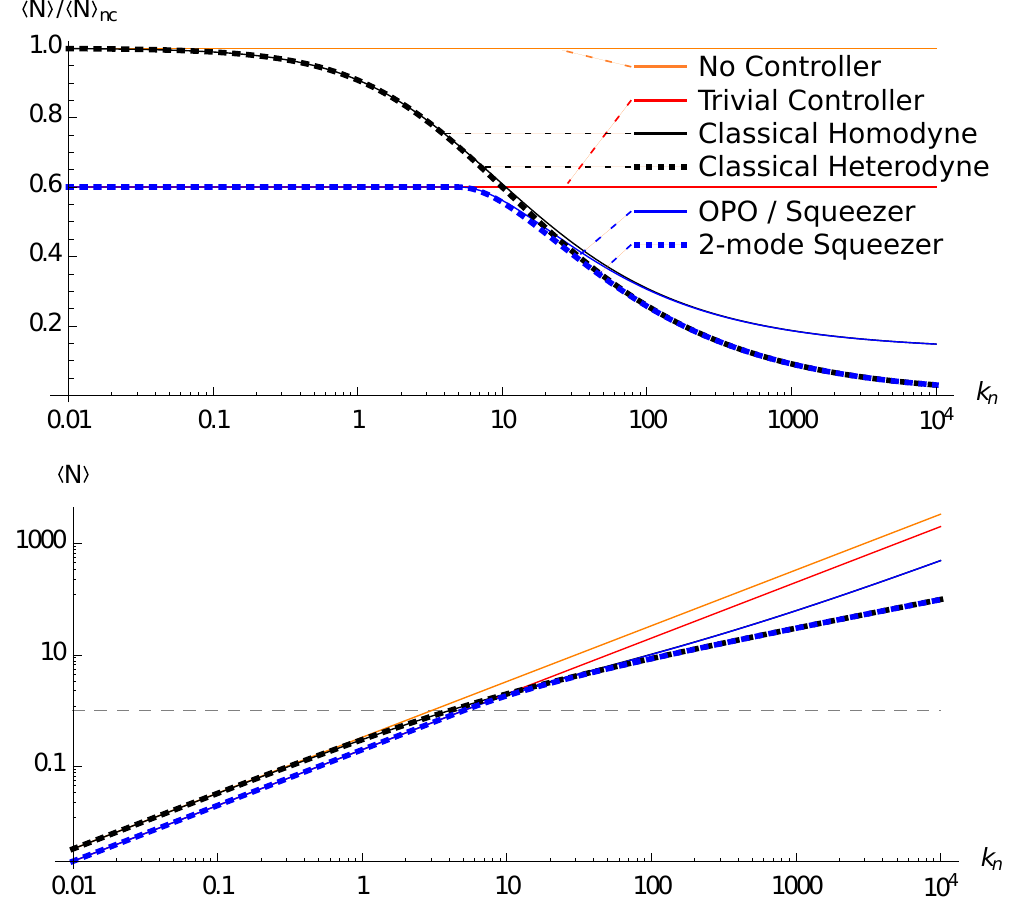}
	\caption{Bottom: Plant Ccavity photon number, as a function of noise strength $k_N$.  The uncontrolled case is shown, as well as the photon number for various control schemes.  Top: Photon number relative to the no-control case.  Smaller is better.}
	\label{fig:f4}
\end{figure}

For a model system with the parameters
\beq
	k_1 = k_2 = k_3 = 0.01, \Delta = 0.1
\eeq
we plot the cost function $\avg{a^\dagger a}$ as a function of noise $k_n$ for the various controller types in Figure \ref{fig:f4}.  The orange line gives the performance of the plant without a controller.  As expected, the photon number rises linearly with the noise power.  It is not hard to show that this matches the analytic result
\beq
	\avg{a^\dagger a}_{\rm nc} = \frac{k_3}{k_1 + k_2 + k_3} k_n \label{eq:cav-nc}
\eeq
that one can derive from the QSDE's.

The trivial controller is simple enough that it also has an analytic solution.  The two mirrors, rather than leaking photons separately, do so constructively so that the leakage {\it amplitudes} (rather than their {\it powers}) add up.  This requires the replacement $k_1+k_2 \rightarrow (\sqrt{k_1} + \sqrt{k_2})^2$ in (\ref{eq:cav-nc}), leading to the following result
\beq
	\avg{a^\dagger a}_{\rm tr} = \frac{k_3}{k_1 + k_2 + k_3 + 2\sqrt{k_1 k_2}} k_n \label{eq:cav-tr}
\eeq
which agrees with the numerical data plotted in Fig. \ref{fig:f4}.

\subsection{Classical Controllers}

More sophisticated are the classical measurement controllers.  The first simply makes a homodyne measurement of the $d\tilde{a}_{1x}$ field.  This signal is fed through a classical circuit which generates an output.  The heterodyne controller is slightly more complicated, and can be modeled as a two-input homodyne measurement controller in the following circuit (using the notation of~\cite{Gough09,QHDLPaper}; see Appendix A):
\beq
	(Hom)_{2-\rm in} \triangleleft (I_1 \boxplus e^{i\pi/2}) \triangleleft BS(\alpha) \label{eq:gj-het}
\eeq
In addition to the homodyne controller's parameters, we can also vary the beamsplitter transmittance.  (Setting the beam-splitter transmission coefficient $\alpha \rightarrow 1$ would send all the light entering controller input $1$ into the $x$-quadrature homodyne detector, so the classical homodyne controller is really a special case of the classical heterodyne controller.)

This example, in particular, illustrates the power of the Gough-James circuit algebra in treating control problems when the controller has a more complex, ``circuit-like'' structure.  Having written code to output the $ABCD$ model for a general $n$-input homodyne controller, it would have been straightforward, albeit tedious, to write additional code for the $n$-input heterodyne controller.  But using the Gough-James circuit algebra allows us to write the $n$-input heterodyne system in terms of a $2n$-input homodyne system, plus some beamsplitters and phase shifters, so we get the heterodyne controller for free.  By breaking the system into smaller components, we can reduce the total amount of work we need to do in quantum control and simulation problems.

There also exist ``analytic'' formulas for LQG-optimal classical controllers in the classical case.  It is not difficult to rewrite Eq.\ (\ref{eq:abcd}) in the standard form for an LQG problem:
\begin{eqnarray}
	dx & = & A_p x dt + B_p du + dw \nonumber \\
	dy & = & C_p x dt + dv \label{eq:class-abcd}
\end{eqnarray}
Here $dy$ is the measurement signal, $du$ is the controller output, and $dw$ and $dv$ are the plant and controller noises, $dw \sim N(0, F_w dt)$, $dv \sim N(0, F_v dt)$.  Unfortunately, in this system the noises are correlated; the vacuum noise $da_1$ acts on both the plant and, after reflection off mirror $k_1$, the controller.  One can define a covariance matrix $M_{ik} = \avg{dw_i dv_k}$ to account for this correlation.

A common trick is to remove the noise correlations by performing a change of variables \cite{Simon06}.  Since $dy - C_p x dt - dv = 0$, we can subtract this quantity from the first line of (\ref{eq:class-abcd}) to find an equivalent equation of motion:
\begin{eqnarray}
	dx & = & A_p x dt + B_p du + dw + M F_v^{-1}(dy - C_p x dt - dv) \nonumber \\
	& = & (A_p - M F_v C_p) x + B_p (du + B_p^{-1}M F_v^{-1}dy) \nonumber \\
	& & + (dw - M F_v^{-1}dy) \nonumber \\
	& = & \tilde{A} x + B_p d\tilde{u} + d\tilde{w}
\end{eqnarray}
Here, the noises $dv$ and $d\tilde{w}$ are uncorrelated.  The controller for this plant will consist of a Kalman filter and a feedback:
\begin{eqnarray}
	d\hat{x} & = & (\tilde{A} - B_p L - K C_p) \hat{x} + K dy \nonumber \\
	d\tilde{u} & = & -L \hat{x} dt
\end{eqnarray}
The Kalman gain and feedback matrices can be obtained by solving the Riccati Equations:
\begin{eqnarray}
	K = \sigma C^{\rm T} F_v^{-1} & & (A\sigma + \sigma A^{\rm T} - \sigma C^{\rm T} F_v^{-1} C \sigma = 0) \nonumber \\
	L = R^{-1} B^{\rm T} \lambda  & & (A^{\rm T}\lambda + \lambda A + Q - \lambda B R^{-1} B \lambda = 0) \nonumber \\
\end{eqnarray}
(Here $Q$ and $R$ are LQR optimization weights for the plant and controller states; we assume $Q\gg R$). For this particular case we optimized the classical controllers numerically, but the results agree with the analytical expression.  When optimizing the measurement controllers, we found that the best controllers always had dynamics that were much faster than the plant timescales. When this happens, the controller's internal dynamics can be {\it adiabatically eliminated} and the controller can be replaced by a simplified ``limit model'' of the original component~\cite{BoutenAd,BoutenAd2,Goug10}.  When a linear component is adiabatically eliminated, its internal variables are removed and its ABCD model is replaced by the input-output relations:
\beq
	d\tilde{a} = (D - C A^{-1} B) da
\eeq
The homodyne controller, adiabatically eliminated, becomes:
\begin{eqnarray}
	d\tilde{a}_x & = & \xi_1 da_x + da_{k1,x} \nonumber \\
	d\tilde{a}_p & = & \xi_2 da_x + da_{k1,p}
\end{eqnarray}
In this device, the signal $da_x$ is measured, amplified by factors $c_1$ and $c_2$, and imprinted onto the output field.  The downside of this measurement is the additional noise $da_{k1}$ that the output accrues.

\begin{figure}[t]
	\centering
	\includegraphics[width=1.00\columnwidth]{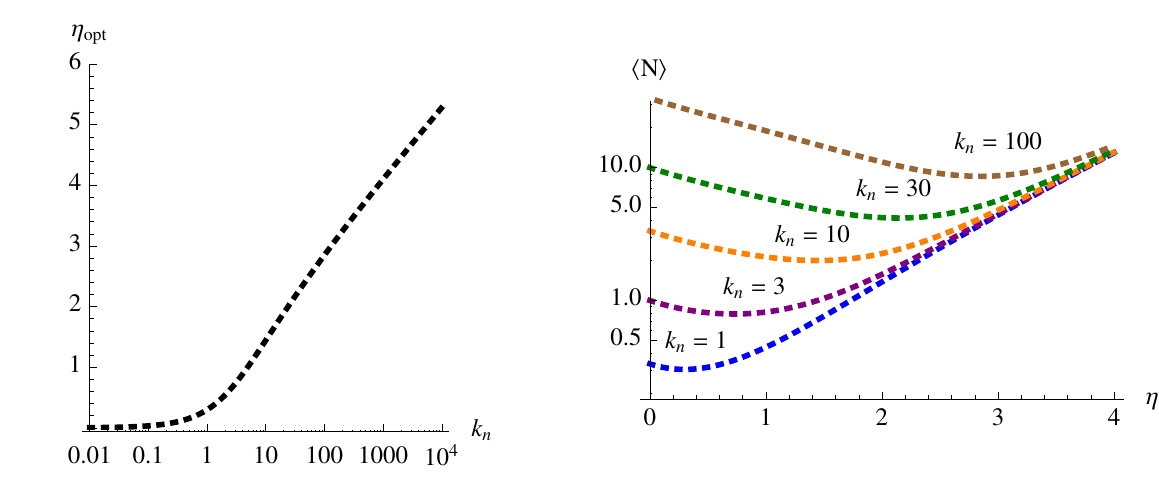}
	\caption{Left: Optimal heterodyne amplification $\eta$ as a function of plant noise.  Right: LQR as a function of controller amplification, for five different noise values.}
	\label{fig:f6b}
\end{figure}

The optimal heterodyne controller uses a 50-50 beamsplitter so we set $\alpha = 1/\sqrt{2}$ in (\ref{eq:gj-het}).  It too has very fast dynamics that can be adiabatically eliminated to give:
\begin{eqnarray}
	d\tilde{a}_x & = & \xi (da_x + da_{k1,x}) + da_{k2,p} \nonumber \\
	d\tilde{a}_p & = & \xi (da_p - da_{k1,p}) + da_{k2,p}
\end{eqnarray}
Or equivalently:
\beq
	d\tilde{A} = \xi (dA + dA_{k1}^\dagger) + dA_{k2}
\eeq
The heterodyne controller amplifies both quadratures, but there is an additional noise due to splitting the beam before measurement, $dA_{k1}$, as well as the measurement noise itself.  The LQR can be computed analytically, and the analytic result agrees with the numerical optimizer.  Setting $\xi = \sinh(\eta)$, we have:
\beq
	\avg{a^\dagger a}_{\rm cl} = \frac{k_2 \sinh^2\eta + k_3 k_n}{k_1 + k_2 + k_3 + 2\sqrt{k_1 k_2} \sinh\eta} \label{eq:lqr-cav-cl}
\eeq
This is plotted in Fig. \ref{fig:f6b}.  As the plant noise increases, so does the controller's optimal amplification.  It does not do well to increase the amplification indefinitely, however, since this also adds noise into the system.  From Fig. \ref{fig:f4}, one can also see that measurement control does well at reducing the photon number for large $k_n$, but in the quantum regime, $k_n \lesssim 1$, it has hardly any effect at all.

\subsection{Coherent Control}

The three coherent controllers of interest are the cavity controller and the two OPO setups, as shown in Figure \ref{fig:f1}.  The optimizer consistently showed that the best cavity controller is in fact the trivial controller (which is the special case of a cavity with mirror transmittivity set to zero).  Because of this, we do not consider empty cavity controllers in this section.  The OPO controllers, on the other hand, have more interesting behavior.

\begin{figure}[t]
	\centering
	\includegraphics[width=1.00\columnwidth]{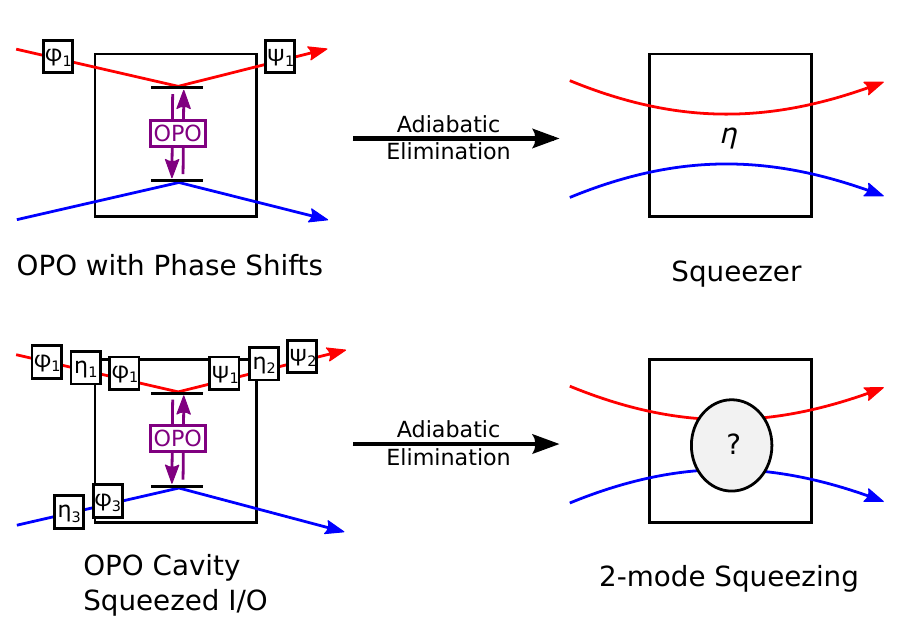}
	\caption{Optical parametric oscillators adiabatically eliminate into ideal squeezers.}
	\label{fig:f5}
\end{figure}

As in the classical case, it was discovered that the best coherent controllers always had dynamics that were much faster than the plant timescales and could be adiabatically eliminated.  A single OPO will adiabatically eliminate to a squeezer with the following input-output relations:
\beq
	d\tilde{a}_x = e^{\eta} da_x,\ \ \ d\tilde{a}_p = e^{-\eta} da_p
\eeq
(up to input and output phase shifts).  An OPO system with squeezed inputs and outputs, which can in principle replicate any 2-port linear quantum system with a single internal degree of freedom \cite{NurdinNetworkSynthesis}, will adiabatically eliminate to arbitrary two-mode squeezer (Fig.\ \ref{fig:f5}).  As far as this control problem is concerned, the {\it best} two-mode squeezer is the linear amplifier, given by the input-output relations:
\begin{eqnarray}
	d\tilde{a}_{1} & = & \cosh(\eta) da_1 + \sinh(\eta) da_2 \nonumber \\
	d\tilde{a}_{2} & = & \sinh(\eta) da_2 + \cosh(\eta) da_2
\end{eqnarray}

Analytic formulas can be derived straightforwardly from the quantum stochastic differential equations.  For the squeezer:
\beq
\avg{a^\dagger a}_{\rm sq} = \frac{\mbox{Re}\left[\left(k_2\sinh^2\eta  + 2 k_n\right) - 2 \frac{k_2\sqrt{k_1k_2}\cosh\eta\sinh^2\eta}{G + 2i \Delta} e^{i \phi}\right]}{\mbox{Re}\left[G - 4k_1k_2\sinh^2\eta/(G + 2i\Delta)\right]} \label{eq:sq-1md}
\eeq
where
\[
	G \equiv k_1 + k_2 + k_3 + 2\sqrt{k_1 k_2}\cosh(\eta) e^{i\phi}
\]
For the linear amplifier:
\beq
	\avg{a^\dagger a}_{\rm 2-sq} = \frac{k_2 \sinh^2\eta + k_3 k_n}{k_1 + k_2 + k_3 + 2\sqrt{k_1 k_2}\cosh\eta} \label{eq:sq-2md}
\eeq
Qualitatively, the results for the heterodyne controller, Eq.\ (\ref{eq:lqr-cav-cl}) and the linear amplifier, Eq.\ (\ref{eq:sq-2md}) look very similar.  Both the heterodyne controller and the linear amplifier reduce the cavity's photon number by amplifying the feedback signal, but also add noise to the system.  For equivalent levels of amplification (compare (\ref{eq:lqr-cav-cl}), substituting $\sinh\eta \rightarrow \cosh\eta$, to (\ref{eq:sq-2md})) the classical controller adds extra noise into the system from the measurement process.  When $k_n$ and $\eta$ are large, this extra noise is negligible, but in the quantum regime where $k_n$ and $\eta$ are $\lesssim 1$, this noise can play a major role in making the linear amplifier outperform the heterodyne controller.

As far as optimization is concerned, Equations (\ref{eq:sq-1md}--\ref{eq:sq-2md}) are simple enough to apply.  Finding the best controller just involves minimizing these functions with respect to $\eta$.  But remember that it was not at all obvious that the best quantum controller should be an adiabatically eliminated squeezer.  This had to be demonstrated by optimizing the general OPO controller, which has many more parameters, and comparing the result to that of the squeezer.  This required a {\it Mathematica} package to quickly convert circuit diagrams to ABCD models, and an efficient optimizer to find the best controller parameters.

Notice that, for large $k_n$, the performance of the two quantum controllers follows the classical performance.  In the classical limit, the OPO / squeezer is amplifying a single quadrature and feeding this back into the plant (with the proper phase shift).  Likewise, the classical controller measures a single quadrature, amplfies that signal and sends this back into the plant.  Thus, the OPO / squeezer is a ``homodyne-like'' controller in the classical limit.  By contrast, the linear amplifier amplifies both modes equally and feeds back the result, making it a ``heterodyne-like'' controller which tracks the performance of the heterodyne controller in the classical limit.

\begin{figure}[t]
	\centering
	\includegraphics[width=1.00\columnwidth]{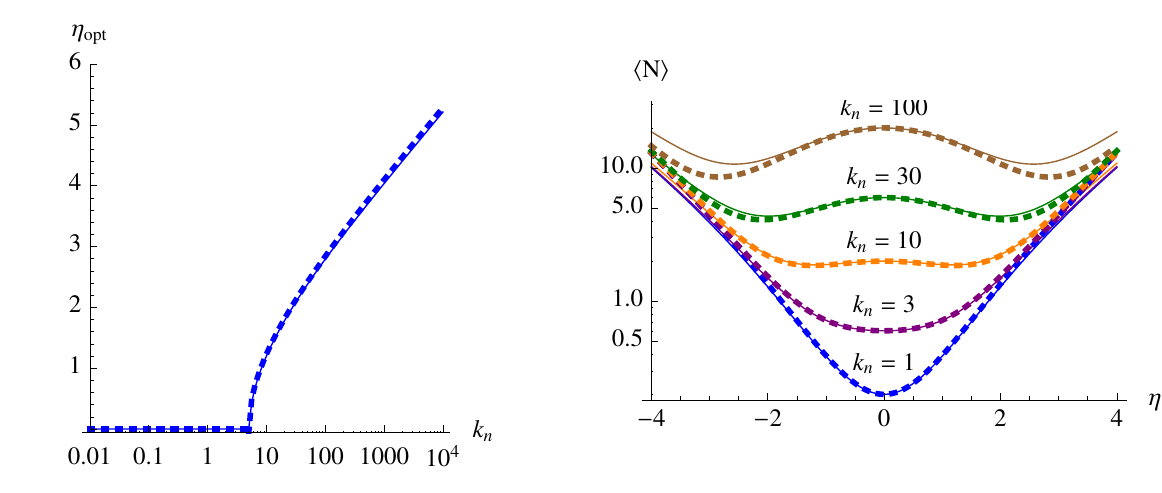}
	\caption{Left: Optimal squeezing for the squeezer (solid) and two-mode squeezer (dotted) controllers.  Right: Performance as a function of squeezing for multiple noise levels.}
	\label{fig:f6}
\end{figure}

However, in the quantum regime, this distinction is lost and both of the quantum controllers track the performance of the trivial controller.  Below a threshold value of
\beq
	k_{n,\rm min} = \frac{k_1(k_1 + k_2 + k_3 + 2\sqrt{k_1 k_2})}{\sqrt{k_1 k_2}k_3}
\eeq
(for this system, $k_{n,\rm min} = 5$), any squeezing will {\it increase} the noise in the cavity, so the optimal value of $\eta$ is zero -- in other words, for $k_n \leq 5$, the best controller is the trivial controller.

As Fig.\ \ref{fig:f6} illustrates, when $k_n > 5$, the best controller has a nonzero amount of squeezing.  We plot the controller performance as a function of squeezing for five different noise levels on the right pane of the figure.  Intuitively, this is a battle between the noise {\it introduced} by squeezing and the noise {\it removed} by constructive interference with the light leaking out mirror $2$.  When $\eta$ is low, the latter dominates.  By increasing the squeezing, we effectively increase the amplitude of the field impinging upon mirror $2$.  Recall that the trivial controller worked by constructive interference between this field and the light leaking out of mirror $2$.  By increasing this field's amplitude, we magnify the effect of this interference; this reduces the overall cavity photon number.  This explains the $\cosh\eta$ term in the denominator of (\ref{eq:sq-2md}).  But a squeezed vacuum carries photons of its own, and some of these photons leak back into the cavity.  If the squeezing is too high, this winds up increasing the photon number, giving rise to the $\sinh^2\eta$ term in (\ref{eq:sq-2md}).  Above the threshold temperature $k_{n,\rm min}$, the ideal $\eta$ lies somewhere between these extremes.

Below the threshold temperature, the cavity photon number is so low that the interference effect never wins out -- squeezing the control field always introduces more photons in the cavity, and the best controller involves looping the output from mirror $1$ into mirror $2$ without squeezing -- the trivial controller.

\section{Optical Feedback Control of a Mechanical Oscillator}

Optomechanical oscillators -- mechanical springs that couple to an optical field via a cavity -- have been a topic of tremendous recent interest in the physics community~\cite{MarquardtGirvin}. A central goal has been to find ways to exploit optomechanical coupling to cool the mechanical oscillator from ambient temperature to its ground state, using optical feedback.

\begin{figure}[b]
	\centering
	\includegraphics[width=1.00\columnwidth]{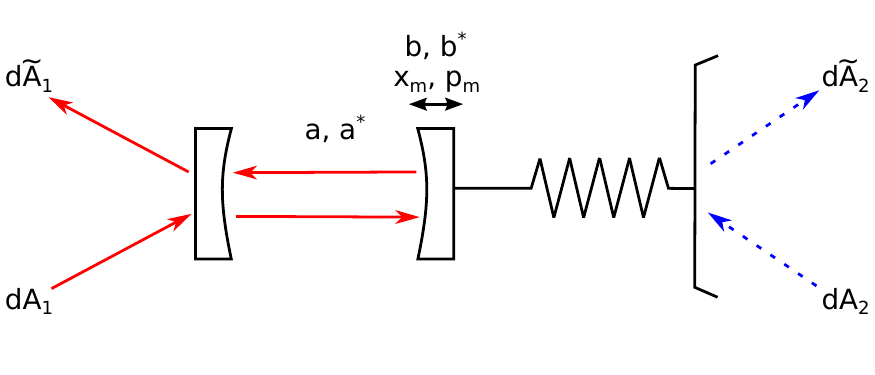}
	\caption{Single cavity with modes $a, a^\dagger$, coupled to a mechanical oscillator with modes $b, b^\dagger$.}
	\label{fig:f7}
\end{figure}

In this section we analyze the optomehcanical oscillator as a coherent control system, with the spring comprising the plant, and with optical probing and feedback.
We optimally cool the oscillator by solving the LQG control problem for the cost function $\avg{b^\dagger b}$, where $b$ is the spring's annihilation operator. While the control setups we consider may appear impractical from an experimental perspective, we will discuss how they can be related to systems that are more realistic to implement.

At the heart of this control problem is the ``adiabatically eliminated cavity,'' depicted in Figure \ref{fig:f7}.  If we go into the rotating frame for the light, this has the SLH model
\beq
	S = 1_{2\times 2},\ \ \ L = \left[\sqrt{\kappa}a, \sqrt{\Omega/Q}b\right],\ \ \
	H = \hbar\Omega b^\dagger b + \eta a^\dagger a x_m \label{eq:slh-mems}
\eeq
where $\Omega$ is the natural spring frequency, $Q$ is the Q-factor, $\kappa$ is the cavity decay parameter, and $m$ is the mirror mass.  See Table \ref{tbl:t1}.

\begin{table}[b]
	\begin{tabular}{r|cl}
		Qty & & Value \\ \hline
		$K_i$ & = & $4\eta r_i/\kappa_i$ \\
		$r_i$ & = & $\sqrt{P_i/\hbar\omega}$ \\
		$\kappa_i$ & = & $t_i c/2l_i$ \\
		$\eta$ & = & $(\omega/l_i)\sqrt{\hbar/2m\Omega}$ \\
		$k_n$ & = & $(1 - e^{-\hbar\Omega/kT})^{-1}$ \\
		$k_m$ & = & $\Omega/Q$ \\
		& &
	\end{tabular}
	\begin{tabular}{p{0.1\columnwidth}|p{0.6\columnwidth}|p{0.3\columnwidth}}
		Qty & Description & Typical Values \\ \hline
		$P_i$ & Laser power in coherent displacement $r_i$, $i = 1, 2$ & $1\,\mu$W--$1\,$mW \\
		$t_i$ & Power transmittance for cavity mirror $i$.  Inversely proportional to finesse. & $10^{-6}$--$10^{-3}$ \\
		$l_i$ & Length of cavity $i$ & $10^{-6}$--$10^{-1}$m \\
		$m$ & Mass of spring-mounted mirror & $10^{-15}$--$10^{-10}$kg \\
		$\Omega$ & Spring oscillation frequency & kHz--GHz \\
		$Q$ & Spring quality factor & $10^{3}$--$10^7$ \\
		$\omega$ & Laser frequency & $2$--$4\times10^{15}$/s
	\end{tabular}
	\caption{Parameters for the optical cavity controller problem.  See, e.g.\ \cite{MarquardtGirvin}}
	\label{tbl:t1}
\end{table}

System (\ref{eq:slh-mems}) is nonlinear by virtue of the interaction term $\eta a^\dagger a X$.  This term is due to the photon pressure of the field in a cavity, which exerts a physical force on the mirror.  In the limit that the light mode $a$ evolves much faster than the mechanical mode $b$, we can adiabatically eliminate the former to give an SLH system of the form:
\beq
	S = \left[\begin{array}{cc} e^{i\phi(x_m - x_{m0};\eta/\kappa)} & 0 \\ 0 & 1\end{array}\right],
	\ \ \ L = \left[0,\sqrt{\Omega/Q}b\right],\ \ \ H = \Omega b^\dagger b \label{eq:el-cav}
\eeq
where
\beq
	\phi(z; \eta/\kappa) = 2\tan^{-1}(2\eta z/\kappa)
\eeq
is the phase shift of the cavity reflected light, as a function of the mirror position (we have absorbed a factor $-1$ in $S$ for convenience).  This is still a highly nonlinear system.  A real optomechanical oscillator is usually driven by a coherent field, and the output that is measured is generally interfered with an equal and opposite field, so as to discern the phase fluctuations on a homodyne detector.  Thus, the real plant system we are interested in is the adiabatically eliminated cavity sandwiched between two coherent displacements.  For a cavity subject to a coherent input of amplitude $r_1$, we write this as:
\beq
	(\mbox{Cav}_1) = L(-r) \triangleleft (\mbox{Cav}) \triangleleft L(r)
\eeq
This has the simple, linear SLH model:
\beq
	S_1 = 1_{2\times 2},\ \ \ L_1 = \left[K_1 x_m, \sqrt{\Omega/Q}b\right],\ \ \
	H_1 = \Omega b^\dagger b \label{eq:sand-cav}
\eeq
with $K_1 = 4\eta r_1/\kappa_1$ is the effective coupling between the spring and the field, which need not be positive or even real.  The $x_m$-coupling to the field $da_1$ gives rise to the following input-output relations:
\begin{eqnarray}
	& & \left\{\begin{array}{rcl}
		(dx_m)_1 & = & 0 \\
		(dp_m)_1 & = & -2K_1 da_{1p}
		\end{array}\right. \nonumber \\
	& & \left\{\begin{array}{rcl}
		d\tilde{a}_{1x} & = & da_{1x} + 2K_1 x_m dt \\
		d\tilde{a}_{1p} & = & da_{1p}
		\end{array}\right.
\end{eqnarray}
The state variable $x_m$ is imprinted on the output $d\tilde{a}_{1x}$, so by measuring the $x$-quadrature of the output field, we can deduce the value of $x_m$; this allows us to use the mirror as a ``measurement'' device, learning information from the output field.  Note that this only works for $d\tilde{a}_{1x}$; no information is imprinted onto the $p$-quadrature of the output.  Conversely, by sending in a particular input $da_{1p}$, we can alter the state of the system; this allows us to use the mirror as a ``feedback'' device.  Note likewise that feedback is not possible via the $da_{1x}$ channel, which does not affect the system.

\begin{figure}[t]
	\centering
	\includegraphics[width=1.00\columnwidth]{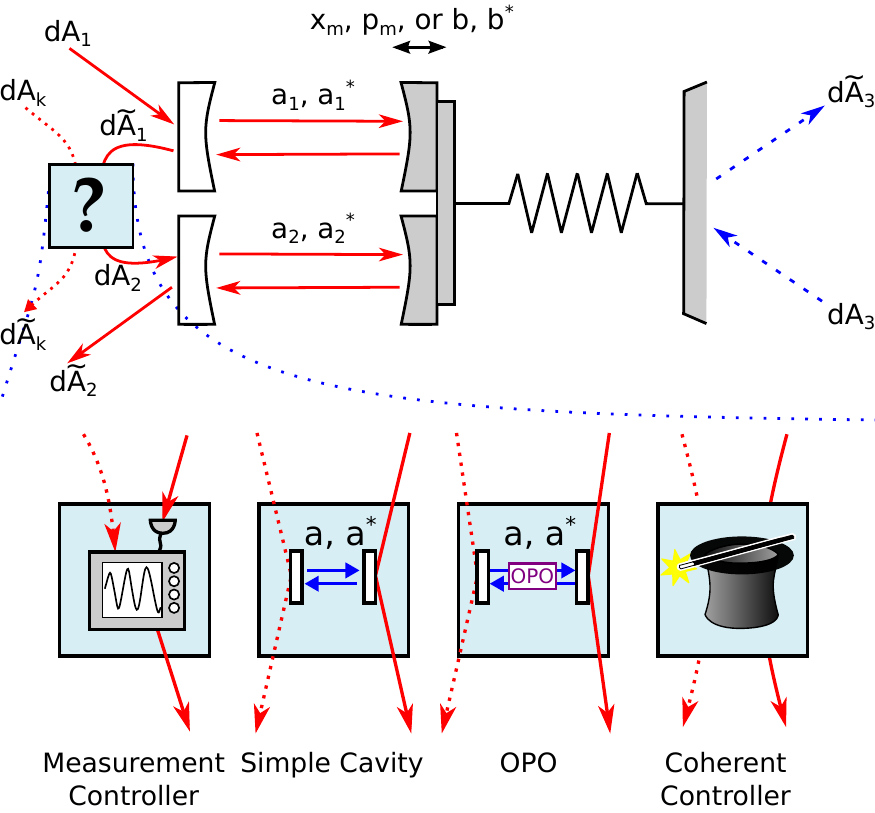}
	\caption{Control-system setup for the mechanical oscillator cooling problem.  Four potential controller designs.}
	\label{fig:f8}
\end{figure}

\subsection{Plant System}

\begin{figure}[t]
	\centering
	\includegraphics[width=0.70\columnwidth]{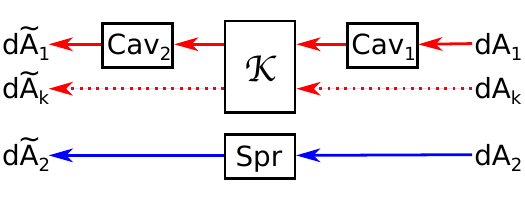}
	\caption{Equivalent view of the plant-controller setup shown in Figure \ref{fig:f8}.  See Eq.\ (\ref{eq:alg-spring}).}
	\label{fig:f9}
\end{figure}

The plant-controller setup is shown in Figure \ref{fig:f8}.  The plant system consists of two (adiabatically eliminated) cavities coupled to the same mirror.  The output from the first cavity, $d\tilde{A}_1$, goes into the controller, and the controller output is fed back into the second cavity input $dA_2$.  Not shown are the two coherent displacements (lasers) putting fields $dA_1$ and $dA_k$ into nonvacuum coherent states.  These coherent fields allow us to replace the cavity with model (\ref{eq:el-cav}) with the linearized model (\ref{eq:sand-cav}).  Since the system is now linear, this becomes an LQG control problem.  The combined plant-controller system can be viewed as a feedback loop from output $d\tilde{A}_1$ to input $dA_2$, or conversely, we can write it as a series product
\beq
	(\mbox{Sys}) = \left[(\mbox{Cav}_2 \boxplus I_1) \triangleleft \mathcal{K} \triangleleft
		(\mbox{Cav}_1 \boxplus I_1)\right] \boxplus (\mbox{Spr}) \label{eq:alg-spring}
\eeq
where (Sys) is the combined system, $\mbox{Cav}_i$ is the $i^{th}$ cavity, with SLH model $(1, \sqrt{k_i}x_m, \_)$, (Spr) gives the spring and phonon couplings, SLH model $(1, \sqrt{k_m} x_m, \Omega b^\dagger b)$, and $\mathcal{K}$ is the controller.  See Figure \ref{fig:f9}.

The controllers we consider here are not unlike those for the simple cavity.  It is not difficult to show using Eq.\ (\ref{eq:alg-spring}) that the trivial controller amounts to no control at all at best, and additional noise at worst.  The classical controller measures the output from mirror cavity 1 and sends an input in to cavity 2, as a function of the controller's internal state.  (Note that we only need to consider a classical controller that measures the $x$ quadrature $d\tilde{a}_{1x}$; $d\tilde{a}_{1p}$ contains no information about the plant's state.)  The simple cavity and OPO cavity coherently process the signal rather than destroying it in a measurement.  Finally, we considered the most general coherent controller, an open quantum system specified by arbitrary $A, B, C, D$ matrices satisfying the realizability relations.  For the LQG problem of minimizing $\avg{b^\dagger b}$, we found optimal controllers in each class for the following plant system:
\begin{eqnarray}
	\Omega & = & 100\ \mbox{(arbitrary units)} \nonumber \\
	k_m & = & 0.01 \nonumber \\
	Q & = & 10000 \nonumber \\
	k_n & = & 10^{-9}\mbox{--}10^9
\end{eqnarray}
In the optimization, we are allowed to vary both the controller parameters {\it and} the couplings $K_1$, $K_2$ to the cavities in (\ref{eq:sand-cav}).  This is because the couplings depend on the input laser powers $P_i$ (in addition to the mirror transmittances $t_i$), which are external quantities (see Table \ref{tbl:t1}) rather than fixed properties of the plant itself.  Here we will operate primarily under the assumption $K_1 = -K_2 \equiv K$; this is a reasonable assumption that avoids classical solutions with divergent controller gain, but we also show that the coherent controllers discussed here outperform the best classical controllers even when this assumption is relaxed.

\begin{figure}[t]
	\centering
	\includegraphics[width=1.00\columnwidth]{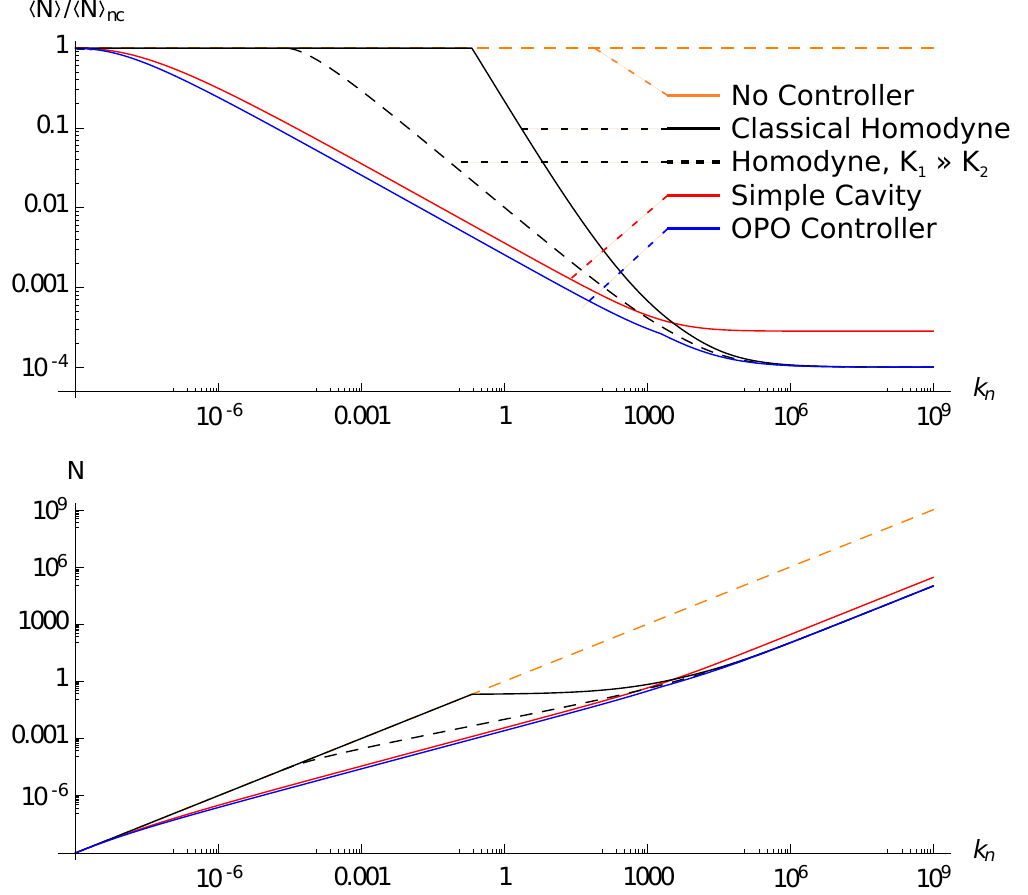}
	\caption{Bottom: Plot of the average phonon number $\avg{N} = \avg{b^\dagger b}$ of the mechanical oscillator for three different control schemes.  Top: Phonon number reduction, relative to no-control case.  The general coherent controller result is not shown, since it overlaps the OPO line, the optimal coherent controller being an OPO cavity.}
	\label{fig:f10}
\end{figure}

Figure \ref{fig:f10} plots the performance of the measurement, simple cavity, and OPO controllers.  For very low ambient temperatures where the noise is weak, the plant is nearly in its ground state to begin with, and none of the controllers can reduce its value.  This differs from the optical cavity.  In the cavity, we used a ``trivial controller'' to cause the light leaking out of mirror $1$ to interfere constructively with the light leaking out mirror $2$, increasing the net dissipation from $k_1+k_2$ to $(\sqrt{k_1} + \sqrt{k_2})^2$.  No such scheme exists in the oscillator because phonons do not ``leak out'' of the system the same way photons leak out of an optical cavity.

At high temperatures, the best classical controller and the OPO controller do equally well, each reducing the phonon number by a factor of exactly $Q = 10000$.  The cavity controller does reasonably well, reducing the phonon number by a factor of about $0.354Q = 3540$.  These results are not very surprising.  The high-temperature limit takes our oscillator into the classical regime, where vacuum noise is negligible and no coherent controller can hope to outperform the best classical controller.

The interesting region lies between these two limits.  Here, there is a sharp cutoff, near $k_n \approx 0.2$, below which the classical measurement controller becomes useless.  As explained below, the classical controller must add noise to the system to make a measurement; below a certain threshold, the gains from control are offset by the noise from measurement.  In this region, the cavity and OPO controllers do significantly better than the classical controller, in some places by a factor of 100--200.

\subsection{The Classical Controller}

\begin{figure}[b]
	\centering
	\includegraphics[width=1.00\columnwidth]{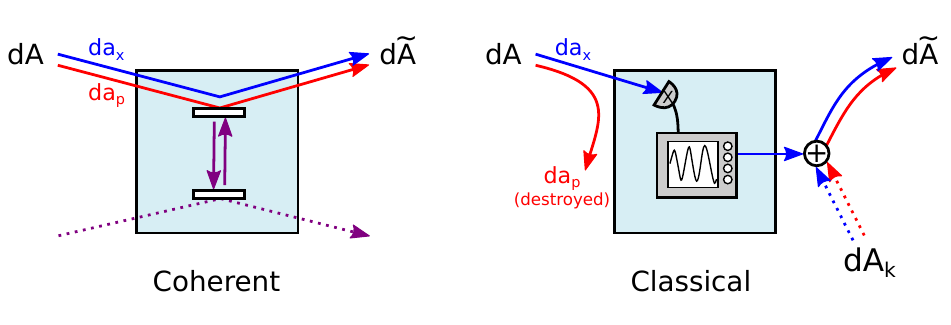}
	\caption{Flow of $x$- and $p$-quadrature signals (blue and red, respectively) in the classical and coherent controllers.}
	\label{fig:f16}
\end{figure}

The classical controller works by measuring the plant output field ($dA$ in Figure \ref{fig:f16}) and inferring the plant's state from this measurement.  From the inferred plant state, the controller applies a feedback signal, which is added to an auxiliary vacuum input $dA_k$ and sent back to the plant.

The plant output contains two quadratures, but only one of them contains information about the system.  Thus, in our classical controller we choose to measure the $x$-quadrature of the output, and necessarily discard the $dA_p$.  This is the optimal control strategy in the classical case because $dA_p$ does not contain any information about the system.  Like any LQG-optimal controller, the classical controller consists of a Kalman filter, which estimates the plant state, plus a feedback element.

The classical controller adds two sources of noise to the plant.  First, by sending a laser through the measurement cavity $(\mbox{Cav}_1)$, it adds {\it measurement noise}, with an amplitude that scales as $O(K)$.  Second, the feedback field $d\tilde{A}$ (with a vacuum noise component due to the auxiliary field $dA_k$) is sent through the controller, adding a {\it feedback noise} of equal magnitude, also $O(K)$.  Both of these factors increase the cavity phonon number by $O(K^2)$, independent of the noise $k_n$.  The control loop will decrease the cavity phonon number by an amount proportional to the present phonon number, which increases with $k_n$.  In the high-$k_n$ limit, the ``control'' term dominates and the coupling $K$ is large.  By contrast, in the low-$k_n$ limit, the ``noise'' term is dominant, and the optimal value of $K$ is small or zero -- no measurement controller can effectively reduce the phonon number, since the noise incurred will more than offset any gains from control.

An important thing to note is the role the $p$-quadrature field $da_p$ plays in this noise budget.  It is true that $da_p$ does not contain any information about the plant state.  But this quadrature still plays an important part, since $da_p$ gives rise to the noise in the measurement cavity, and $d\tilde{a}_p$ gives rise to the noise in the feedback cavity.  Because $da_p$ and $d\tilde{a}_p$ are independent (the former being destroyed in the $da_x$ measurement), their noises add up.  The beauty of coherent control is that we can process the $x$-field without destroying $da_p$ and the measurement and feedback noises become correlated.  If done right, they cancel each other out.

\begin{figure}[b]
	\centering
	\includegraphics[width=1.00\columnwidth]{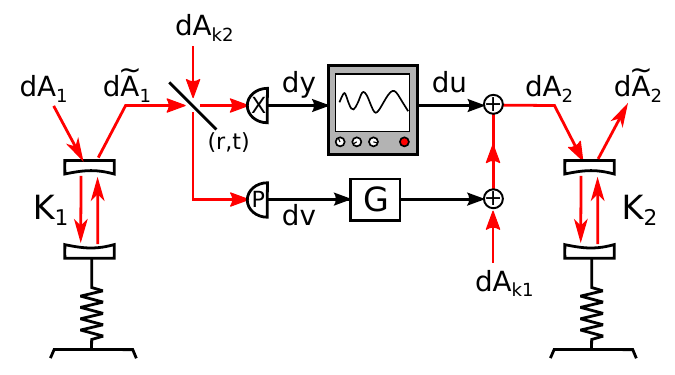}
	\caption{Heterodyne-based measurement controllers, which measure both quadratures of the beam by splitting it, do not not outperform the best homodyne controller for this system.}
	\label{fig:f17}
\end{figure}

If we are free to relax the $K_1 = -K_2$ assumption, then the classical controller does somewhat better (dashed line in Figure \ref{fig:f10}), but still underperforms the coherent schemes discussed below.  When $K_1 \neq K_2$, the optimal classical controller tends to have $K_2 \ll K_1$, which greatly suppresses the measurement noise.  To compensate for this disparity, the controller must have a large classical gain.

It might be thought that a heterodyne-based control scheme, like that in Figure \ref{fig:f17} could perform better than the best homodyne controller.  After all, the homodyne controller is just a special case of the heterodyne controller, where the beamsplitter has a transmissivity of 100\%.  Moreover, one might imagine using a heterodyne scheme to cycle part of the $d\tilde{a}_p$ quadrature back into the plant, canceling out part of the measurement noise with the feedback noise.  However, we find numerically that the most general heterodyne controller does not perform any better -- either with $K_1 = K_2$ or not.  The extra noise added from splitting the beam outweighs any of the benefits of the control scheme.

\subsection{Simple Cavity Controller}

\begin{figure}[b]
	\centering
	\includegraphics[width=1.00\columnwidth]{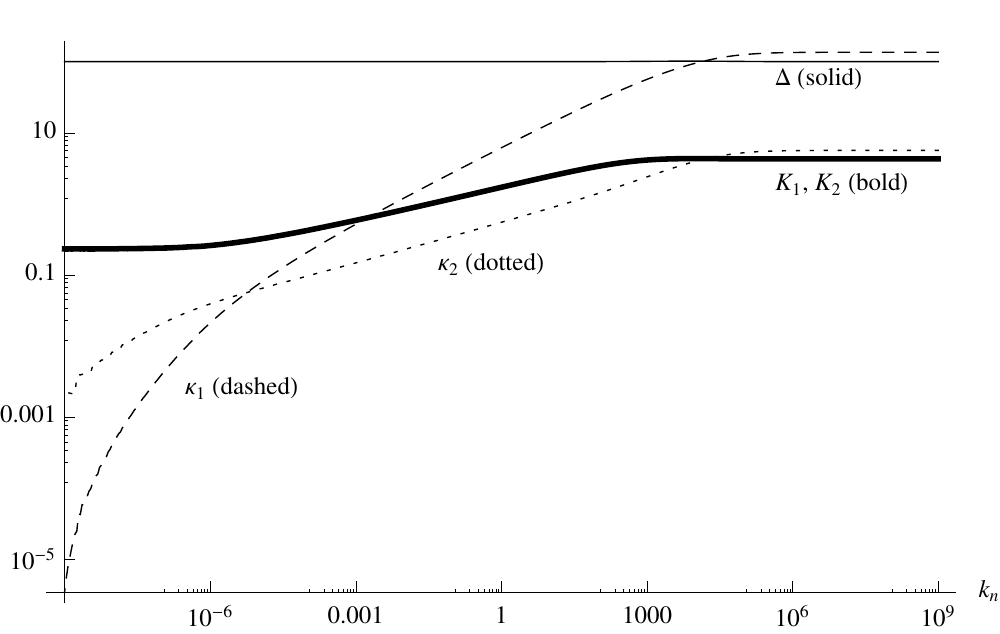}
	\caption{Parameters of the optimal simple cavity controller, as a function of noise strength.}
	\label{fig:f11}
\end{figure}

An empty optical cavity with two input / output ports has the following SLH model:
\beq
	S = 1_{2\times 2},\ \ \ L = \left[\sqrt{\kappa_1}a, \sqrt{\kappa_2}a\right],\ \ \
	H = \Delta a^\dagger a
\eeq
Here the $\kappa$'s are mirror decay parameters and $\Delta$ is the detuning of the cavity.  The QSDEs for the cavity are easy to derive:
\begin{eqnarray}
	da & = & (-i\Delta - \kappa/2)a\,dt + \sqrt{\kappa_1} d\tilde{A}_1 + \sqrt{\kappa_2} d\tilde{A}_2 \nonumber \\
	d\tilde{A}_i & = & dA_i + \sqrt{\kappa_1}a\,dt
\end{eqnarray}
Remember that, in addition to the controller parameters, we can vary the input coherent fields, which allows us to vary the plant's  $x_m$-coupling $K$.  The laser field impinging on cavity 1 adds shot noise to the mirror; in this setup, since $K_1 = -K_2 \equiv K$, the shot noise from cavity 1 will exactly cancel the shot noise from cavity 2 (if we let $K_1$ and $K_2$ vary freely, we find that the optimal controller has $K_1 = -K_2$).  As a consequence, the cavity controller has {\it neither} measurement nor feedback noise.

The optimal detuning and couplings are plotted in Figure \ref{fig:f11}.  Not surprisingly, as the noise on the mirror is increased, the couplings $K_{1,2}$ and $\kappa_{1,2}$ increase as well.  The detuning $\Delta$, which shows no dependence on the noise power, always remaining at a constant value $\Delta \approx \Omega = 100$ for this system, making the cavity controller setup analogous to two coupled harmonic oscillators, one mechanical and the other optical \cite{Botter12}.  Absent the couplings, the quadratures $x = a + a^\dagger, p = (a - a^\dagger)/i$ would evolve just as the mirror variables $x_m, p_m$.

This can also be interpreted as a form of sideband cooling.  The detuning $\Delta \approx \Omega$ indicates that our control system is being driven by laser light at a frequency $\omega_{\rm cav} - \Omega$, where $\omega_{\rm cav}$ is the cavity resonance frequency.  The plant-controller coupling serves to convert photons of frequency $\omega_{\rm cav} - \Omega$ to photons of frequency $\omega_{\rm cav}$, cooling the oscillator.  At high temperatures, we need a large cooling rate to counter the noise; this is achieved by using a cavity with a broad bandwidth $\kappa$, so that both $\omega_{\rm cav} - \Omega$ and $\omega_{\rm cav}$ photons are interact effectively with the cavity.  Conversely, at low temperatures, we need to work in the resolved sideband limit $\kappa \ll \Omega$ to suppress quantum fluctuations of the radiation-pressure force \cite{Miao10, Marquardt07, WilsonRae07}.

\begin{figure}[b]
	\centering
	\includegraphics[width=1.00\columnwidth]{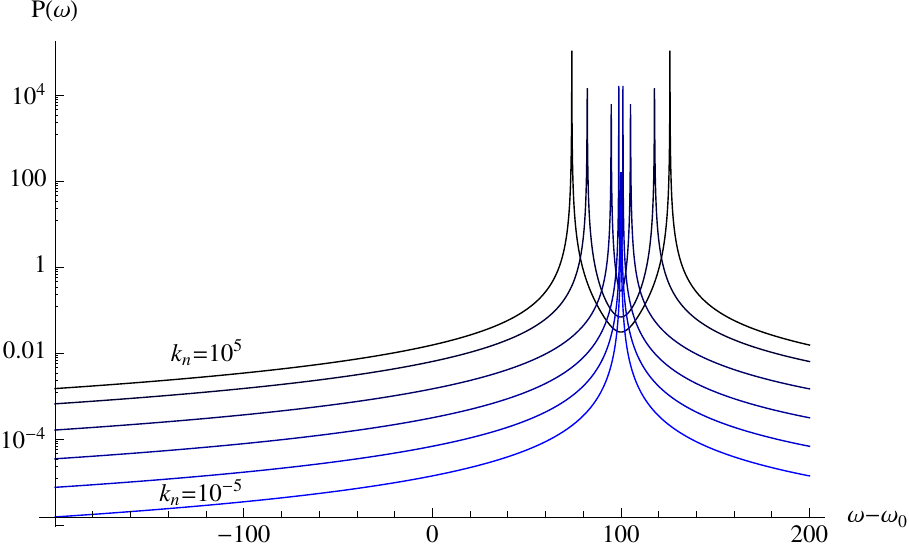}
	\caption{Calculated output spectrum of light exiting the optimal simple cavity controller.  Six values of $k_n$ are plotted, $10^5$ (darkest), $10^3$, $10^1$, $10^{-1}$, $10^{-3}$, and $10^{-5}$ (lightest).}
	\label{fig:f19}
\end{figure}

The effects of this cooling are made manifest on the output power spectrum of the photon channel $\tilde{P}_1(\omega) = \tilde{A}_1(\omega)^\dagger \tilde{A}_1(\omega)$, where $\tilde{A}_1(\omega)$ is the Fourier transform of the stochastic process $d\tilde{A}_1(t)$.  In the frequency domain, the relevant QSDEs for the combined plant-cavity system are
\begin{eqnarray}
	-i\omega a & = & \Bigl[\bigl(-i\Delta - (\kappa_1+\kappa_2)/2\bigr)a + \sqrt{\kappa_1} K (b + b^\dagger)\Bigr] \nonumber \\
	& & + i\omega \sqrt{\kappa_1} A_1 + i\omega \sqrt{\kappa_2} A_2 \nonumber \\
	-i\omega b & = & \left[\left(-i\Omega - \Omega/2Q\right)b - \sqrt{\kappa_1} K (a - a^\dagger)\right] + i\omega \sqrt{\Omega/Q} A_3 \nonumber \\
	-i\omega \tilde{A}_1 & = & -i\omega A_1 + \sqrt{\kappa_1}a
\end{eqnarray}
This power spectrum is plotted in Figure \ref{fig:f19}.  As the exiting light is blue-detuned, it reduces the phonon number in the oscillator, driving it towards the ground state.  For small $k_n$, when the plant and controller are weakly coupled, there is a single sideband corresponding to the plant's oscillation frequency $\Omega$.  When $k_n$ is large, the plant and controller become strongly coupled and the combined system resonates at two different frequencies, one larger than $\Omega$ and one smaller.  This is the origin of the sideband splitting in the figure.

The system can also be understood as a form of coherent Kalman filtering.  Recall that the optimal classical controller works as a Kalman filter, reproducing the state of the plant by measuring one of its outputs.  The cost we paid for the Kalman filtering was additional noise added to the system.  The cavity controller can also be thought of as a Kalman filter, but one that preserves the coherence of the input signal $dA$.  From a quantum mechanical standpoint, in the classical controller, the $p$-quadrature $da_p$ is essentially discarded after the measurement.  In the cavity controller, the field retains its coherent properties and the $da_p$ coming out is the same as $da_p$ going in.  This makes the noises in the measurement and feedback cavities correlated.  In the present setup, they exactly cancel out.  This cancellation of the measurement noise is what gives the coherent cavity controller its superior performance, particularly in the low phonon-number regime.

Measurement sensing experiments \cite{TsengCaves}, particularly in the context of LIGO \cite{Arcizet06}, show similar improvements, but for a different performance metric.  This suggests that LQG control is far from the only problem to benefit from this noise cancellation and coherent feedback; similar gains should be expected in all types of control problems when the plant operates in the quantum regime.

\subsection{OPO Cavity Controller}

\begin{figure}[b]
	\centering
	\includegraphics[width=1.00\columnwidth]{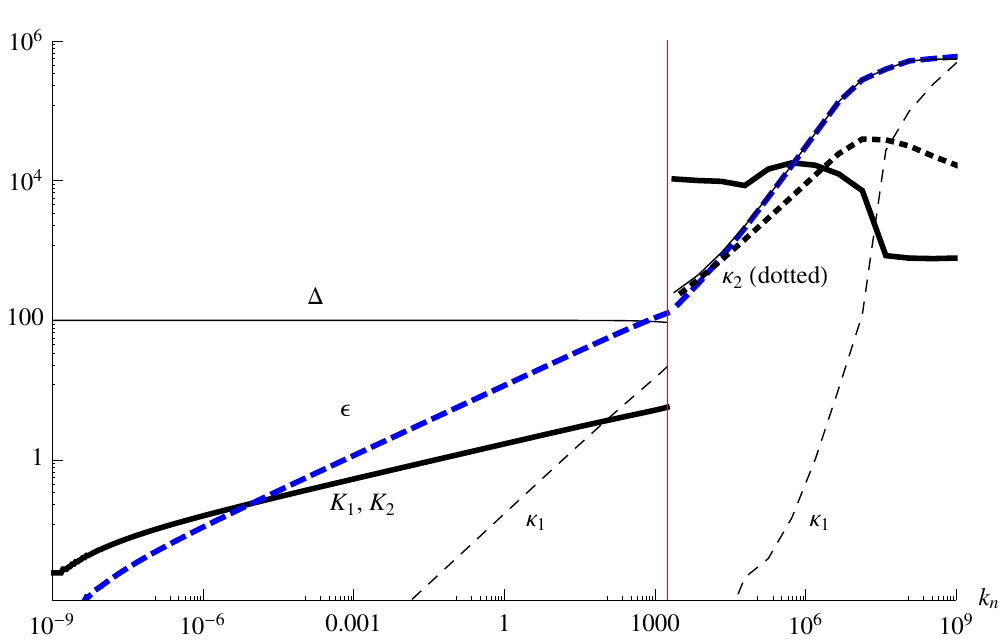}
	\caption{Parameters of the optimal OPO cavity controller, as a function of noise strength.}
	\label{fig:f12}
\end{figure}

Recall from Eq.\ (\ref{eq:opo-slh}) that the OPO has the following SLH model:
\begin{eqnarray}
	& & S = 1_{2\times 2},\ \ L = \left[\sqrt{\kappa_1} a,\ \sqrt{\kappa_2} a\right], \nonumber \\
	& & H = \frac{1}{4} x^{\rm T} \left[\begin{array}{cc} \Delta - \mbox{Im}(\epsilon) & \mbox{Re}(\epsilon) \\ \mbox{Re}(\epsilon) & \Delta + \mbox{Im}(\epsilon) \end{array}\right]x \nonumber \\
	& & \ \ = \Delta a^\dagger a + \frac{\epsilon^* a^2 - \epsilon (a^\dagger)^2}{2i} \label{eq:opo-slh2}
\end{eqnarray}
For fullest generality, the OPO controller is placed between two phase shifters, so the actual controller is $e^{i\phi_1} \triangleleft (\mbox{OPO}) \triangleleft e^{i\phi_2}$.  Between the controller, the phase shifters and the couplings $K_{1,2}$, there are nine free parameters in this LQG problem.  The best OPO controller parameters, found using the optimization code, are plotted in Figure \ref{fig:f12}.  As with the cavity controller, the best OPO controller has $K_1 = -K_2$.

For $k_n \lesssim 1800$, the OPO behaves much like the simple cavity.  Its detuning is close to $\Omega$, the coupling $K_1 = -K_2$ increases with $k_n$, and the mirror losses $\kappa_1, \kappa_2$, while small, increase with increasing noise ($\kappa_2$ is too small to be seen on this plot).  For the most part, $\epsilon \ll \Delta$ and the OPO squeezing is only a perturbation on the dynamics of an empty cavity.

At $k_n \approx 1800$, this changes suddenly.  This happens because the OPO controller has two local minima.  Below $k_n \approx 1800$, the empty cavity-like local minimum is smaller, but above this threshold, a new minimum dominates.  In this regime, the coupling $K$ is much stronger than before and the mirrors $\kappa_1, \kappa_2$ are much more lossy.

The OPO controller appears to be the best coherent controller one can make for this system.  We ran the optimizer for general coherent controller, subject to no constraints other than the realizability conditions (\ref{eq:prc}).  At no point did we find a coherent controller that outperformed the OPO for this system.  This in mind, ths discontinuity at $k_n$ can be better understood.  As the best relizable controller, the OPO must do at least as well as both the simple cavity and the classical controller.  For weak noise, the simple cavity outperforms the classical controller, so we expect the OPO to look more like a simple cavity.  For strong noise, the classical controller does better, so we expect the OPO to look more like a classical controller, inasmuch as this is possible.  There is no reason to assume that the transition between the two {\it must} be smooth.  It may be marked with bifurcation points, as in Figure \ref{fig:f6} for the cavity control problem, or it may occur with a discontinuity in the parameters.  What happens for a general plant / controller system will depend on the landscape of the cost function, and in particular, the behavior of local minima.

\subsection{More Realistic Control Systems}

\begin{figure}[t]
	\centering
	\includegraphics[width=1.00\columnwidth]{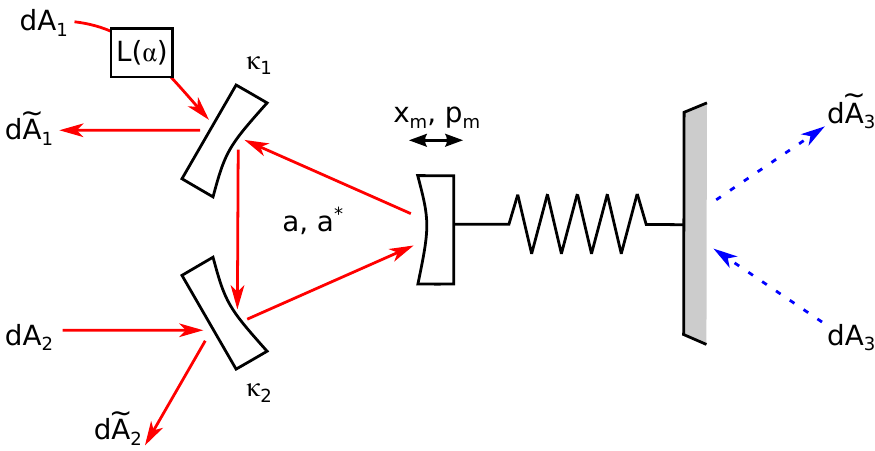}
	\caption{Model for a {\it non}-adiabatically eliminated cavity.}
	\label{fig:f13}
\end{figure}

The control systems discussed above can be implemented in principle, but they require two separate mirrors and two separate cavities to be coupled to the same mechanical oscillator, which may prove difficult to build in a laboratory.  Fortunately, one can show that for the cavity controller and the OPO controller, equivalent systems can be realized using a non-adiabatically eliminated cavity with one of its mirrors on a spring.

First, the simple cavity controller.  Recall from (\ref{eq:alg-spring}) that the cavity controller system can be modeled as
\beq
	\left[(\mbox{Cav}_2 \boxplus I_1) \triangleleft (\mbox{Cav}) \triangleleft (\mbox{Cav}_1 \boxplus I_1)\right] \boxplus (\mbox{Spr})
\eeq
which has the SLH model
\begin{eqnarray}
	& & S = 1_{2\times 2},\ \ \ L = \left[\sqrt{\kappa_1}a, \sqrt{\kappa_2}a, \sqrt{\kappa_m}b\right] \nonumber \\
	& & H = \omega_c a^\dagger a + \Omega b^\dagger b + \sqrt{\kappa_1}K_1 x_m p_c \label{eq:slh-spr1}
\end{eqnarray}

\begin{figure}[t]
	\centering
	\includegraphics[width=1.00\columnwidth]{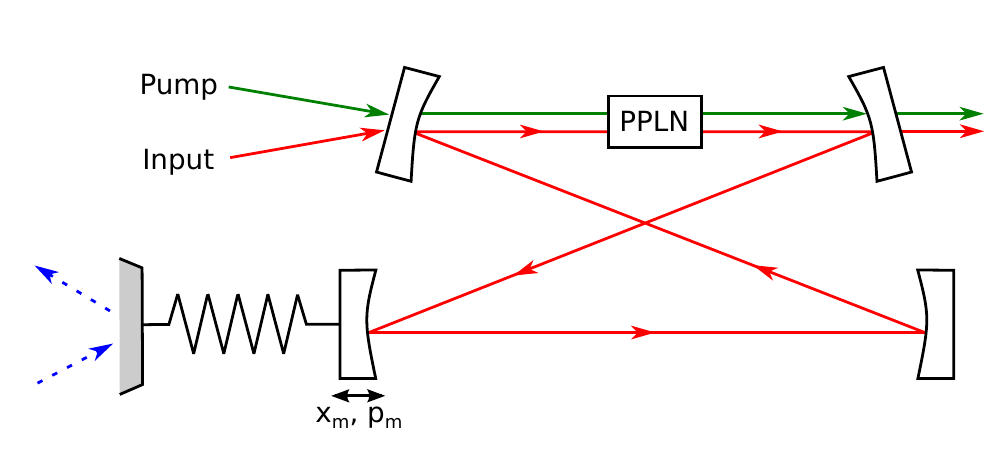}
	\caption{Model for a {\it non}-adiabatically eliminated OPO cavity with a spring mirror.}
	\label{fig:f14}
\end{figure}

Now consider a system, depicted in Figure \ref{fig:f13}, consisting of a {\it non}-adiabatically eliminated cavity with one of its mirrors attached to a spring.  This has the SLH model:
\begin{eqnarray}
	& & S = 1_{3\times 3},\ \ \ L = \left[\sqrt{\kappa_1}a, \sqrt{\kappa_2}a, \sqrt{k_m} b\right] \nonumber \\
	& & H = \Delta_0 a^\dagger a + \Omega b^\dagger b + \eta a^\dagger a x_m
\end{eqnarray}
A laser $L(\alpha)$ sends a coherent input into mirror $1$, giving the system $\mbox{Cav} \triangleleft (L(\alpha) \boxplus I_2)$.  Of course, the internal dynamics do not depend on anything downstream of the system, so we can just as well use $(L(\alpha') \boxplus I_2) \triangleleft \mbox{Cav} \triangleleft (L(\alpha) \boxplus I_2)$, for any $\alpha'$.  Making substitutions $a \rightarrow a - a_0, b \rightarrow b - b_0$ to center around the equilibrium point, the SLH model becomes:
\begin{eqnarray}
	& & S = 1_{3\times 3},\ \ \ L = \left[\sqrt{\kappa_1}a, \sqrt{\kappa_2}a, \sqrt{k_m} b\right] \nonumber \\
	& & H = \Delta a^\dagger a + \Omega b^\dagger b + \frac{\eta|\alpha|\sqrt{\kappa_1}}{\Delta^2 + (\kappa/2)^2} x_m x_c + \eta a^\dagger a x_m \nonumber \\
	& & \label{eq:slh-spr2}
\end{eqnarray}
Ignoring the nonlinear term, this is almost identical to (\ref{eq:slh-spr1}).  One can convert the $x_m x_c$ term to an $x_m p_c$ term with a canonical transformation, and the coefficients can be matched by varying $\alpha$.  Thus the systems in (\ref{eq:slh-spr1}) and (\ref{eq:slh-spr2}) are equivalent, and the ``simple cavity controller'' can be realized in the lab using a single cavity with a mirror attached to a spring.  Cooling an oscillator in this setup has been realized experimentally, though it was not interpreted as a control system \cite{Giga06,Arci06,Klec06}.

The OPO controller is just like the cavity controller, but the Hamiltonian has an additional squeezing term; see (\ref{eq:opo-slh}).  The same procedure can be applied to show that the OPO plant-controller system is equivalent to a (non-adiabatically eliminated) OPO cavity with a spring mirror, as shown in Figure \ref{fig:f14}.

\subsection{Quantum Refrigerator Analogy}

\begin{figure}[t]
	\centering
	\includegraphics[width=1.00\columnwidth]{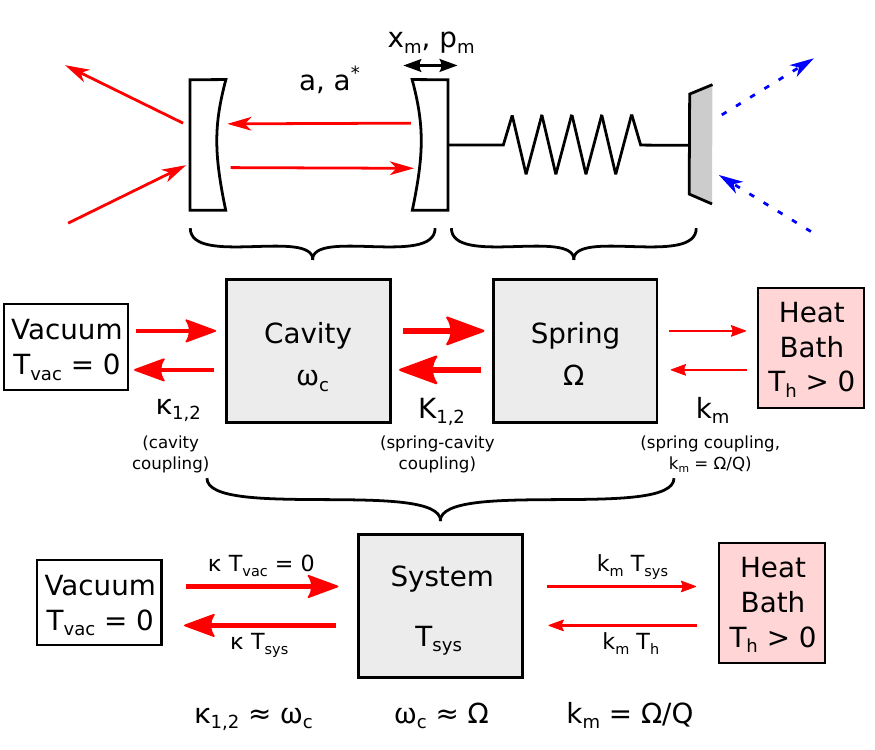}
	\caption{Coherent control problem represented as two coupled thermodynamic systems.}
	\label{fig:f15}
\end{figure}

One thing we notice from the optimal controller performance is that, in the strong-noise limit, the optimal controllers -- classical, OPO, cavity -- all reduce the spring phonon number by a factor of about $Q = \Omega / k_m$.  The classical and OPO controllers reduce it by exactly $Q$, while the cavity controller only reduces it by a factor $Q / 2.83$.  This factor-of-$Q$ reduction can be understood by viewing the plant and the controller as thermodynamic systems.

Figure \ref{fig:f15} illustrates our point.  Starting with a cavity with a spring mirror, we separate the system into the cavity, which oscillates at a frequency $\omega_c$, and the spring, which oscillates at a frequency $\Omega$.  Each system has its own coupling to the environment.  The cavity couples to a vacuum-state environment ($T = 0$) with coupling strengths $\kappa_1, \kappa_2$, the spring, couples to a heat bath with $T_h > 0$ with strength $k_m$, and a spring-cavity coupling $K_1 = -K_2$ couples the two modes.

If the spring and cavity oscillate at about the same frequency and the spring-cavity coupling is strong compared to the other two, then the ``temperature'' of the spring will be roughly equal to the ``temperature'' of the cavity.  We denote this temperature $T_{\rm sys}$.  One expects the combined system to be in thermal steady-state with both the heat bath and vacuum inputs and outputs; this gives us the energy balance equation:
\beq
	k_m T_h = k_m T_{\rm sys} + \kappa T_{\rm sys}
\eeq
where $\kappa = \kappa_1 + \kappa_2 \sim \omega_c \sim \Omega$, and $k_m = \Omega / Q$.  Solving for the system's steady-state temperature,
\beq
	T_{\rm sys} = \frac{k_m T_h}{\kappa + k_m} \sim \frac{T_h}{Q}
\eeq
From general arguments, we can therefore expect that most good controllers will reduce the spring phonon number by a factor of about $Q$, but that no controller will do significantly better.  Note that, since this argument is based on thermodynamic assumptions that are only approximately valid here, the factor-of-$Q$ reduction is only approximate, and only holds in ths classical limit.  These classical results, unsurprisingly, break down in the quantum regime because, among other things, the effects of vacuum noise inputs become important.

\section{Conclusions}

In this paper, we have studied the coherent-feedback cooling of linear quantum systems from an LQG control perspective.  The systems were modeled using the SLH framework and the Gough-James circuit algebra, which allow arbitrarily large circuits be constructed in a straightforward and systematic manner.  The evolution of the system was studied using QSDEs, the open-system analogue to the Heisenberg Equations. We wrote {\it Mathematica} scripts based on the QHDL/M framework to model quantum LQG control systems, and designed algorithms to optimize a controller's parameters for a given setup.

For any LQG control problem, there is always a quantum controller that does at least as well as the optimal classical controller. In the quantum regime, when excitation number in the plant is of order unity, we have shown that the best quantum controller can do better -- in some cases, significantly so.  Two systems -- the optical cavity and the optomechanical oscillator -- were studied in detail.  For the former, modest gains were found using coherent control in the low-photon-number regime.  For the latter, the gains were much larger.

One could imagine extending these results to look at non-quadratic cost functions in linear control systems. Indeed, some work has already been done on this matter, focusing on using coherent feedback to maximize the squeezing in a cavity mode \cite{Iida11}. Taking a control theory perspective may also provide insight into minimizing the noise in optomechanical sensors. In addition, the understanding the superior performance of coherent feedback in linear systems may provide important clues for the design of quantum controllers for nonlinear systems such as optical switches or error correcting codes.

\begin{acknowledgements}
This work has been supported by the NSF (PHY-1005386), AFOSR (FA9550-11-1-0238) and DARPA-MTO (N66001-11-1-4106). Ryan Hamerly is supported by the NSF GRFP and a Stanford Graduate Fellowship. We thank Nikolas Tezak, Gopal Sarma, Dmitri Pavlichin and Orion Crisafulli for useful discussions.
\end{acknowledgements}

\appendix
\section{SLH, QSDE, ABCD models} \label{sec:slh-app}

An open quantum system with $n$ bosonic, Markovian input-output channels, can be represented as a triple \cite{Gough08, Gough09}:
\beq
	(S, L, H)
\eeq
where $S$ is a unitary $n \times n$ operator-valued matrix, $L$ is an $n$-component operator-valued vector, and $H$ is Hermitian.

\begin{figure}[t]
	\centering
	\includegraphics[width=1.00\columnwidth]{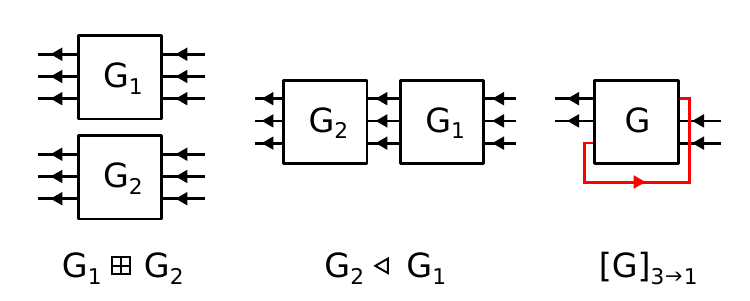}
	\caption{The concatenation, series, and feedback products are the three basic operations of the Gough-James circuit algebra.}
	\label{fig:f18}
\end{figure}

A quantum circuit is built up by connecting together components of this form.  Any such a circuit can be written in terms of a {\it circuit algebra} of concatenation, series and feedback products, shown in Figure \ref{fig:f18}.

The concatenation product joins two systems without connecting any of the input/output ports.  If $G_1 = (S_1, L_1, H_1)$ and $G_2 = (S_2, L_2, H_2)$ then the concatenation is given by:
\beq
	G_1 \boxplus G_2 = \left( \left[\begin{array}{cc} S_1 & 0 \\ 0 & S_2 \end{array}\right],\ \left[\begin{array}{c} L_1 \\ L_2 \end{array}\right],\ H_1 + H_2\right)
\eeq
The series product feeds ths outputs of one system into the inputs of another, and has the following SLH model:
\beq
	G_2 \triangleleft G_1 = \left(S_2 S_1,\ L_2 + S_2 L_1,\ H_1 + H_2 + \mbox{Im}(L_2^\dagger S_2 L_1)\right)
\eeq
The feedback product can $[G]_{i\rightarrow j}$ corresponds to taking output $i$ of $G$ and feeding it back into input $j$.  This can also be represented with an SLH model.  See Ref.\ \cite{QHDLPaper}.

The Quantum Stochastic Differential Equations (QSDEs) are Heisenberg-picture equations of motion for open quantum systems.  They relate the evolution of the internal state variables (denoted $X$) and the output fields $d\tilde{A}_i$ and gauge processes $d\tilde{\Lambda}_{ij}$ to the inputs $dA_i$, $d\Lambda_{ij}$, where the inputs are vacuum quantum Wiener processes.  For a given SLH model, the equations are \cite{Gough09}:
\begin{eqnarray}
	dX & = & \left[-i\comm{X}{H} + \frac{1}{2}\left(L_i^\dagger [X, L_i] + [L_i^\dagger, X] L_i\right)\right] dt \nonumber \\
	& & + dA_i^\dagger S_{ij}^\dagger [X, L_j] + \nonumber \\
	& & [L_j^\dagger, X] S_{ji} dA_i
	+ \left(S_{ik}^\dagger X S_{kj} - X \delta_{ij}\right) d\Lambda_{ij} \\
	d\tilde{A}_i & = & S_{ij} dA_j + L_i dt \\
	d\tilde{\Lambda}_{ij} & = & S^*_{ik} d\Lambda_{kl} S^{\rm T}_{lj} + S^*_{ik}dA^\dagger_k L_j
	+ L^\dagger_{i} dA_k S^{\rm T}_{kj} + L^\dagger_i L_j \nonumber \\
\end{eqnarray}
Likewise, the Master Equation is the Schr\"{o}dinger-picture equation of motion for an open quantum system.  It gives the evolution of the system's density operator:
\beq
	\frac{d\rho}{dt} = i[\rho, H] + \left(L_i \rho L_i^\dagger - \frac{1}{2} L_i^\dagger L_i \rho - \frac{1}{2} \rho L_i^\dagger L_i\right)
\eeq

A linear system has a quadratic Hamiltonian and linear environment couplings.  It takes the following SLH model:
\begin{eqnarray}
	S_{ij} & = & S_{ij} \nonumber \\
	L_i & = & \Lambda_{i}x + \lambda_i \nonumber \\
	H & = & \frac{1}{2} x^{\rm T} R x + r^{\rm T}x
\end{eqnarray}
From these, the ABCD matrices take the form:
\beq
\begin{array}{rclrcl}
	A & = & 2\Theta\left(R + \frac{1}{4}\tilde{\Lambda}^{\rm T} J \tilde{\Lambda}\right),\ \ \  &
	B & = & \Theta \tilde{\Lambda}^{\rm T} J \tilde{S}, \\
	C & = & \tilde{\Lambda}, &
	D & = & \tilde{S}, \\
	a & = & 2\Theta\left(r + \frac{1}{4}\tilde{\Lambda}^{\rm T} J \tilde{\lambda}\right), &
	c & = & \tilde{\lambda}
\end{array}
\eeq
We form matrices $\tilde{S}$ and $\tilde{\Lambda}$, and vector $\tilde{\lambda}$ by stacking $S$, $\Lambda$, and $\lambda$:
\begin{eqnarray}
	\tilde{S}_{ab} & = & 2M^\dagger \left[\begin{array}{cc} S_{ab} & 0 \\ 0 & S_{ab}^* \end{array}\right] \nonumber \\
	\tilde{\Lambda}_a & = & 2M^\dagger \left[\begin{array}{c} \Lambda_{a} \\ \Lambda_{a}^* \end{array}\right] \nonumber \\
	\tilde{\lambda}_a & = & 2M^\dagger \left[\begin{array}{c} \lambda_{a} \\ \lambda_{a}^* \end{array}\right]
\end{eqnarray}
where
\beq
	J_{2n\times 2n} = I_n \otimes \left[\begin{array}{cc} 0 & 1 \\ -1 & 0 \end{array}\right]
\eeq
is the canonical antisymmetric matrix of dimension $2n$ (written above as $J$, where the dimension is inferred), and
\beq
	M_{2n\times 2n} = I_n \otimes \frac{1}{2}\left[\begin{array}{cc} 1 & i \\ 1 & -i\end{array}\right]
\eeq
The matrix $\tilde{S}$ is made from the blocks $\tilde{S}_{ab}$ above, and likewise for $\tilde{\Lambda}$ and $\tilde{\lambda}$.

These are similar to the formulas used in \cite{JamesNurdinPetersen08}, the difference being that our input and output fields $da$ are written in the quadrature basis, $(da_x, da_p)$ rather than $(da, da^\dagger)$.  This keeps all of our inputs and outputs Hermitian, and also makes all of our matrices real.

\end{document}